# Competitive ligand binding kinetics to linear polymers

Juan P.G. Villaluenga[1,*], David Brunete[1], Francisco Javier Cao-García[1,2,**]

1. Departamento de Estructura de la Materia, Física Térmica y Electrónica, Universidad Complutense de Madrid, Plaza de Ciencias, 1, 28040 Madrid, Spain
2. Instituto Madrileño de Estudios Avanzados en Nanociencia, IMDEA Nanociencia, Calle Faraday, 9, 28049 Madrid, Spain.

 * E-mail: jpgarcia@ucm.es

 ** E-mail: francao@ucm.es

**Abstract**

Different types of ligands compete in binding to polymers with different consequences for the physical and chemical properties of the resulting complex. Here, we derive a general kinetic model for the competitive binding kinetics of different types of ligands to a linear polymer, using the McGhee and von Hippel detailed binding site counting procedure. The derived model allows the description of the competitive binding process in terms of the size of the ligand, binding and release rates, and cooperativity parameters. We illustrate the implications of the general theory showing the equations for the competitive binding of two ligands. The size of the ligand, given by the number of monomers occluded, is shown to have a great impact in competitive binding. Ligands requiring a large available gap for binding are strongly inhibited by smaller ligands. Ligand size then has a leading role compared to binding affinity or cooperativity. For ligands that can bind in different modes (i.e., different number of monomers), this implies that they are more effective in covering or passivating the polymer in lower modes, if the different modes have similar binding energies.

# I. Introduction

Ligands compete to bind to long polymers (*in vitro* and *in vivo*) and the result determines the physical and chemical properties of the resulting complex [1]–[7]. The kinetics of binding processes and their equilibrium state is known to have pharmacological implications [8]–[10], with large efforts devoted to its characterization [8], [11]. In particular, the unspecific binding of ligands to single and double-stranded DNA is known to play a key role in DNA replication and repair [5], [12]–[19]. Particularly, during DNA replication, single-stranded DNA-binding proteins (SSBs) wrap single-stranded DNA (ssDNA) to protect it from degradation and prevent secondary structure formation [3], [14], [20], [21]. As shown for the Escherichia coli SSB (EcoSSB), and Human mitochondrial SSB (HmtSSB), SSBs may bind ssDNA in multiple modes that differ in the number of subunits that contact DNA. For example, EcoSSB may bind ssDNA in four modes, which can be denoted $(SSB)_{17}$, $(SSB)_{35}$, $(SSB)_{56}$ and $(SSB)_{65}$, where the subscripts indicate the average number of ssDNA nucleotides occluded per SSB protein [3], [6], [21]–[24]. Recently, the binding equilibrium and kinetics of SSBs to ssDNA have been studied using optical tweezers [3], [6]. These studies call for accurate models to account for competition between different binding modes and to understand which of the modes is more effective in protecting the ssDNA.

Simplified kinetic models have been proposed to the description of single and multiple ligand modes or types of ligands [25]. However, these simplified models have been shown to give inaccurate prediction for medium and high coverage, even in the single ligand case [26]. This calls for a proper development of the ligand binding positions counting for the equilibrium [27]–[29] and the kinetics [26]. The theory has also been extended for the case where the single ligand is cooperative [30]–[33] The models have also been extended to the equilibrium case for multiple or competitive ligand binding case [34]–[39]. This calls for extending the models to develop kinetic models with detailed binding position counting for the multiple ligands binding case.

Here, we develop a model of competitive ligand binding kinetics to polymers. The model is obtained extending to the multiple ligands case a previous single ligand binding kinetic model [33]. The polymer is modelled as a linear array with uniformly spaced binding sites. As ligands can bind to multiple sites, an appropriate counting of possible binding positions is required, which is done following the McGhee and von Hippel procedure [27], [29]. The model provides the kinetics and the equilibrium states. It gives the fraction of the polymer sites covered by each ligand at any time and at the final equilibrium state.

Here, we consider that cooperativity arises from an enhancement of the binding of ligands to the polymer. Recently, we derived a kinetic approach considering that cooperativity can affect the binding and release rates [33]. Positive cooperativity emerges by an enhancement of the binding or by an inhibition of the release. Negative cooperativity arises as an inhibition of the binding or an enhancement of the release. The results presented here for competitive binding kinetics could be extended to include cooperativity due to inhibition of the release following the procedures presented in Ref. [33].

In Section II, we extend a cooperative ligand binding kinetics model to the case where a mixture of distinct types of ligands bind to the same polymer, using the McGhee and von Hippel procedure. In Section III, we illustrate the implications of the kinetic model describing the ligand binding of two different ligand types. We analyse the competitive binding between two types of ligands in terms of the size of the ligands, their kinetic rate constants, and their cooperativity. Finally, Section IV discusses the results.

## II. Multimode ligand binding to a linear polymer

We derive an approach for the binding of several types (or modes) of ligands to a linear polymer using the procedure proposed by McGhee and von Hippel to count the possible binding sites. Note that the approach is applied for describing non-cooperative binding and cooperative binding to nearby sites.

### II.A. Definition of coverage, conditional probabilities, gap probability and cooperativity parameter

The polymer is modelled as a lattice and the ligand as a single entity binding a fixed number of polymer units. The polymer is represented by a linear array of $N$ identical repeated units, which can be named as monomers or polymer residues. A ligand molecule is assumed to bind to the polymer and to cover $m$ consecutive units (i.e., make inaccessible to another ligand). A free ligand binding site consists of any $m$ consecutive free polymer units. See Fig. 1. The polymer is assumed to be infinitely long, implying $N \gg m$.

### II.A.1. Coverage

We consider a system with several modes (or types) of binding to a polymer. The fraction of polymer monomers covered by the ligand of type $i$ is given by the coverage

$$c_i = \frac{n_i \cdot m_i}{N}, \tag{1}$$

where $n_i$ is the number of ligands of type $i$ bound to the polymer, and $m_i$ is the number of polymer units covered by the ligand of type $i$. In the present approach, a monomer can only exist in two states, free and bound to the ligand. The total coverage of the polymer is

$$c = \sum_{i=1}^{k} c_i, \tag{2}$$

when there are *k* several types of ligands.

### II.A.2. Conditional probabilities

We will describe the states of the polymer residues: free or bound, the probability of this states, and the relations between neighbouring probabilities through conditional probabilities. (We restrict ourselves here only to cooperativity with the adjacent ligand.) The following notation is adopted. Any bound ligand can be divided into $m_i$ sections, each one corresponding to the underlying polymer residue; we number these sections as 1, 2,…, $m_i$ from left to right. (See Fig. 1, Top) Thus, we have $1 + \sum_{i=1}^{k} m_i$ distinguishable types of monomers: a free residue, labelled $f$; a residue under the number 1 or left end of a bound ligand of type $i$, labelled $b_1^i$; and so on from $b_2^i$ up to $b_{m_i}^i$, the latter representing the right end of a bound ligand of type $i$. We can thus denote the conditional probabilities as a sequence of two such types. For example, $(ff)$ is the probability, given a free residue, that another free residue lies to the immediate right; $(fb_1^i)$ is the probability, given a free residue, that the left end of a bound ligand of type $i$ lies to the immediate right; $(b_{m_i}^i f)$ is the probability, given the right end of a bound ligand of type $i$, that a free residue lies to the immediate right (see Fig. 1, Top).

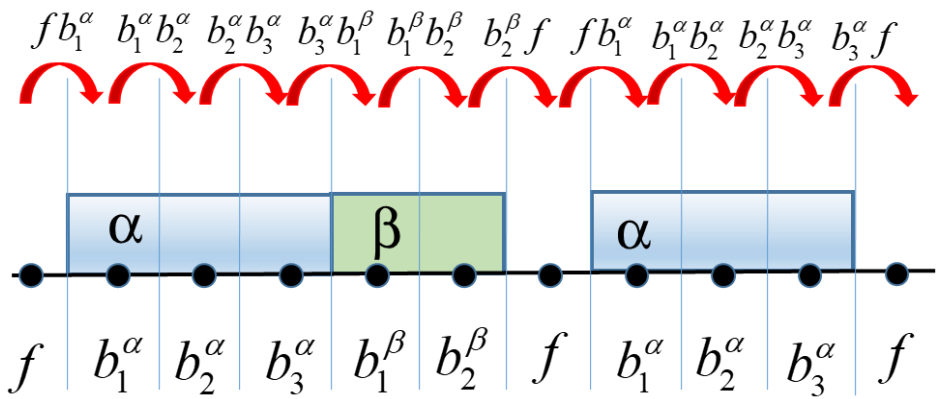

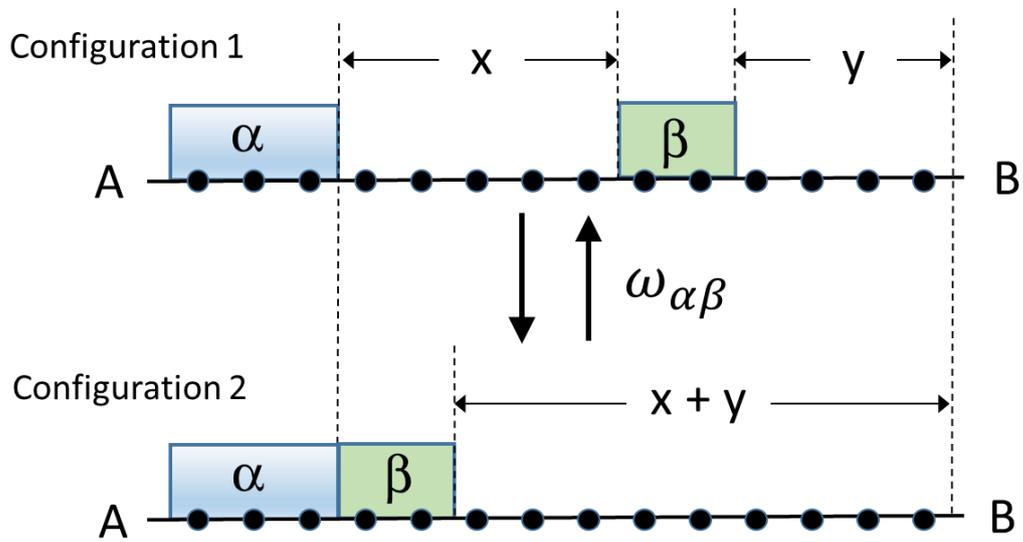

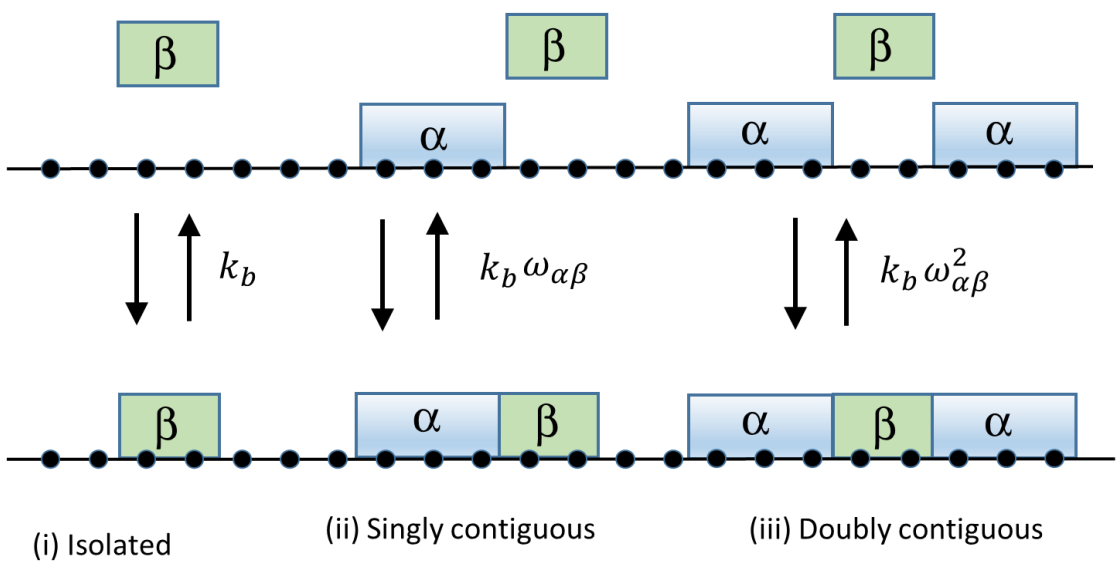

*Figure 1: (Top) Notation for monomer states and conditional probabilities. (Middle) Comparison of two configuration, cooperativity increases the probability of the second one by an amount $\omega_{\alpha\beta}$ (cooperativity parameter for neighbouring ligands of types $\alpha$ and $\beta$). (Bottom) Comparison of effective equilibrium constant for the binding of an isolated, singly, and doubly contiguous ligand, respectively.*

In principle, there are $\left(1 + \sum_{i=1}^{k} m_i\right)^2$ different conditional probabilities that can be expressed by this notation, but the physical fact that all $m_i$ sections of a ligand are contiguous, makes many of these conditional probabilities are either zero or one. The following conditional probabilities are zero: $(fb_i^l)$ for $i \neq 1$; $(b_i^l b_j^l)$ for $j \neq i + 1$; and $(b_i^l b_j^h)$ for $l \neq h$, when $i \neq m_l$ or $j \neq 1$. The conditional probabilities for consecutive units of a bound ligand are one, e.g., $(b_1^l b_2^l) = 1$, $(b_2^l b_3^l) = 1$, etc.

The relevant (non-trivial) conditional probabilities are the $(ff)$, $(fb_1^i)$, $(b_{m_i}^i f)$, and $(b_{m_l}^l b_1^i)$; which are a total of $(1+k)^2$. Between them there are several consistency relations. There are only two types of polymer sites which can possibly lie to the immediate right of a free site: either another free site or the first part of a bound ligand, giving the following relation:

$$(ff) + \sum_{i=1}^{k} \left(fb_1^i\right) = 1. \tag{3}$$

Only a free site or the left end of a second bound ligand can lie to the immediate right of the right end of a bound ligand, implying

$$\left(b_{m_l}^l f\right) + \sum_{i=1}^{k} \left(b_{m_l}^l b_1^i\right) = 1. \tag{4}$$

### II.A.2.a. Non-cooperative case

For non-cooperative ligands, by symmetry, the probability is not dependent on the arrangement of two contiguous bound ligands when they are different ($i \neq m_l$)

$$\left(b_{m_l}^l b_1^i\right) = \left(b_{m_i}^i b_1^l\right). \tag{5}$$

For non-cooperative ligands, if one considers the polymer site immediately to the right of one selected at random, this second site also, by definition, is selected at random and thus also has a probability of $(1-c)$ of being free. There are only two ways in which the two-step random selection can result in a second free site. Either the first site chosen is free (a $f$ site chosen with probability $(1-c)$) and has a free site to its right (conditional probability $(ff)$); or the first site chosen is the right end of a bound ligand (a $b_{m_i}^i$ site chosen with probability $c_i/m_i$) and has a free site to its right (conditional probability $(b_{m_i}^i f)$). Since the overall probability that the second site is free must be independent of the method of random selection, we obtain

$$(1-c)(ff) + \sum_{i=1}^{k} \frac{c_i}{m_i}\left(b_{m_i}^i f\right) = (1-c). \tag{6}$$

### II.A.3. Gap probability

We can now compute the probability of having a gap of a given size, $g$. A gap is a set of $g$ consecutive free ligand binding sites (free lattice residues). Ligands cannot bind to gaps smaller than their size, $g < m_i$. To count the number of free sites in a gap, we start at the right end of one bound ligand, and count by proceeding to the right, one site at a time, until we reach the left end of the next bound ligand. This procedure allows to express the probability that a given gap has a certain size as a product of conditional probabilities of residues states. The probability that any particular gap, between a bound ligand of type $\alpha$ and a bound ligand of type $\beta$, is exactly $g$ free polymer units long is given by

$$\left(P_g\right)_{\alpha\beta} = (b_{m_\alpha}^\alpha f)(ff)^{g-1}\left(fb_1^\beta\right), \tag{7}$$

where $\left(b_{m_\alpha}^\alpha f\right)$ is the probability, having selected the right end of a bound ligand of type $\alpha$, that the polymer site to the immediate right is free; $(ff)$ is the probability, given a free site, that there is a second free site to the immediate right; and $\left(fb_1^\beta\right)$ is the probability, given a free site, that the left end of a bound ligand of type $\beta$ lies to the immediate right.

### II.A.4. Cooperativity parameter

The cooperativity parameter is defined as the ratio of the probabilities of the two configurations represented in Fig. 1 (Middle). Ligand locations at the left (A) and right (B) are the same for both configurations. The difference between the two configurations is that in the second configuration the two represented ligands are at neighbouring locations and have an additional ligand-ligand interaction. This additional ligand-ligand interaction makes Configuration 2 more probable when it is attractive, positive cooperativity. The probability of Configuration 1 is

$$\wp_1 = \wp_A (b^\alpha_{m_\alpha} f)(ff)^{x-1}\left(fb^\beta_1\right)\left(b^\beta_1 b^\beta_2\right)\cdots\left(b^\beta_{m_\beta} f\right)(ff)^{y-1}\wp_B \ . \tag{8}$$

while the probability of Configuration 2 is

$$\wp_2 = \wp_A \left(b^\alpha_{m_\alpha} b^\beta_1\right)\left(b^\beta_1 b^\beta_2\right)\cdots\left(b^\beta_{m_\beta} f\right)(ff)^{x+y-1}\wp_B \ , \tag{9}$$

Thus, the cooperativity parameter, defined as the ratio, becomes

$$\omega_{\alpha\beta} = \frac{\wp_2}{\wp_1} = \frac{\left(b^\alpha_{m_\alpha} b^\beta_1\right)(ff)}{\left(b^\alpha_{m_\alpha} f\right)\left(fb^\beta_1\right)} \ . \tag{10}$$

For $\omega_{\alpha\beta} > 1$, the ligands α and β attract each other, and the binding is positively cooperative; for $\omega_{\alpha\beta} = 1$, the binding is non-cooperative; for $\omega_{\alpha\beta} < 1$, the ligands repel each other, and the binding is negatively cooperative.

**II.B. Derivation of the average number of free binding sites per polymer**

The interaction between ligands is allowed between nearest neighbours when they are bound without intervening free polymer residues. This restriction results in three different types of ligand binding sites: (i) an isolated site to which a ligand binds with a binding constant, $k_b$; (ii) a singly contiguous site to which a ligand binds with binding constant $k_b\,\omega_{\alpha\beta}$; and (iii) a doubly contiguous site to which a ligand binds with binding constant $k_b\,\omega^2_{\alpha\beta}$. The cooperativity parameter $\omega_{\alpha\beta}$ is the equilibrium constant for the process of moving a bound ligand from an isolated site to a singly contiguous site, or from a singly contiguous site to a doubly contiguous site. (See Fig. 1, Bottom).

The number of free ligand binding sites on a naked polymer is $N - m_i + 1$. The number of potential ligand-binding sites eliminated by binding one ligand can range from 1, if the ligand binds into a gap exactly $m_i$ residues long, up to $2m_i - 1$ if it binds to a naked polymer. For a gap $g$ residues long between adjacent bound ligands, the number of binding sites is $g - m_i + 1$ if $g \geq m_i$, but zero if $g < m_i$. The probability of a gap being $g$ free polymer sites long is denoted by $P_g$, and the probability of a gap being $g$ free polymer sites long between a bound ligand of type α and a bound ligand of type β, is denoted by $(P_g)_{\alpha\beta}$. The relation between both probabilities is given by

$$P_g = \sum_{\alpha=1}^{k}\sum_{\beta=1}^{k}(P_g)_{\alpha\beta}. \qquad (11)$$

where $(P_g)_{\alpha\beta}$ is given by Eq. (7).

We next obtain expressions for the average number of free binding sites per gap and per polymer.

### II.B.1. Non-cooperative case

When the binding is non-cooperative ($\omega_{lh} = 1$), the average number of free binding sites per gap, $\bar{s}_\lambda$, is given by

$$\bar{s}_\lambda = \sum_{\alpha=1}^{k}\sum_{\beta=1}^{k}\sum_{g=m_\lambda}^{N}(g - m_\lambda + 1)\frac{c_\alpha}{c}(P_g)_{\alpha\beta}. \qquad (12)$$

The equation (12) can be evaluated, and we obtain that (see Appendix for derivation)

$$\bar{s}_\lambda = \left[\sum_{\alpha=1}^{k}\frac{c_\alpha}{c}\cdot(b_{m_\alpha}^\alpha f)\right]\frac{(ff)^{m_\lambda-1}}{(1-(ff))^2}\left[\sum_{\beta=1}^{k}(fb_1^\beta)\right] = \frac{(ff)^{m_\lambda}}{1-(ff)}. \qquad (13)$$

### II.B.2. Cooperative case

When the binding is cooperative ($\omega_{lh} \neq 1$), as there are three different types of ligand binding sites, we distinguish several cases. For $g < m_l$, there are no free binding sites. For $g = m_\lambda$, there is one doubly binding site, thus

$$(\bar{s}_{dc})_{\alpha\beta}^\lambda = (P_{m_\lambda})_{\alpha\beta} = (b_{m_\alpha}^\alpha f)(ff)^{m_\lambda-1}(fb_1^\beta), \qquad (14)$$

where $\bar{s}_{dc}$ is the average number of free doubly contiguous binding sites. Next, in the polymer, by summing over the different combinations weighted by the cooperativities (see Appendix for derivation), we obtain that

$$(\bar{s}_{dc})^\lambda = \sum_{\alpha=1}^{k}\sum_{\beta=1}^{k}(\bar{s}_{dc})^\lambda_{\alpha\beta}\omega_{\lambda\alpha}\omega_{\lambda\beta} =$$
$$= \left[\sum_{\alpha=1}^{k}\frac{c_\alpha}{c}(b^\alpha_{m_\alpha}f)\omega_{\lambda\alpha}\right](ff)^{m_\lambda-1}\left[\sum_{\beta=1}^{k}(fb^\beta_1)\omega_{\lambda\beta}\right]. \quad (15)$$

For $g \geq m_\lambda + 1$, there are two singly contiguous binding site, thus the sum turns out to be (see Appendix for derivation)

$$(\bar{s}_{sc})^\lambda_{\alpha\beta} = 2\sum_{g=m_\lambda+1}^{\infty}(P_g)_{\alpha\beta} = 2\frac{c_\alpha}{c}(b^\alpha_{m_\alpha}f)\frac{(ff)^{m_\lambda}}{1-(ff)}(fb^\beta_1), \quad (16)$$

where $\bar{s}_{sc}$ is the average number of free singly contiguous binding sites. In the polymer, by summing over the different combinations weighting with the cooperativities, it follows that (see Appendix for derivation)

$$(\bar{s}_{sc})^\lambda = \sum_{\alpha=1}^{k}\sum_{\beta=1}^{k}(\bar{s}_{sc})^\lambda_{\alpha\beta}\frac{\omega_{\alpha\lambda}+\omega_{\lambda\beta}}{2} =$$
$$= \sum_{\alpha=1}^{k}\sum_{\beta=1}^{k}\frac{c_\alpha}{c}(b^\alpha_{m_\alpha}f)\frac{(ff)^{m_\lambda}}{1-(ff)}(fb^\beta_1)(\omega_{\alpha\lambda}+\omega_{\lambda\beta}). \quad (17)$$

For $g \geq m_\lambda + 2$, there are $(g - m_\lambda - 1)$ isolated binding sites per gap, thus the sum becomes (see Appendix for derivation)

$$(\bar{s}_i)^\lambda = \sum_{\alpha=1}^{k}\sum_{\beta=1}^{k}\sum_{g=m_\lambda+2}^{\infty}(g-m_\lambda-1)(P_g)_{\alpha\beta} =$$
$$= \left[\sum_{\alpha=1}^{k}\frac{c_\alpha}{c}(b^\alpha_{m_\alpha}f)\right]\frac{(ff)^{m_\lambda+1}}{(1-(ff))^2}\left[\sum_{\beta=1}^{k}(fb^\beta_1)\right] \quad (18)$$

Finally, by using Eqs. (15)-(18), the average number of free binding sites in the polymer is written as:

$$\bar{s}_\lambda = (\bar{s}_{dc})^\lambda + (\bar{s}_{sc})^\lambda + (\bar{s}_i)^\lambda =$$
$$= \left\{\sum_{\alpha=1}^{k}\frac{c_\alpha}{c}(b^\alpha_{m_\alpha}f)\left[\omega_{\alpha\lambda}+\frac{(ff)}{1-(ff)}\right]\right\}(ff)^{m_\lambda-1}\left\{\sum_{\beta=1}^{k}(fb^\beta_1)\left[\omega_{\lambda\beta}+\frac{(ff)}{1-(ff)}\right]\right\}. \quad (19)$$

For the non-cooperative case ($\omega_{lh} = 1$), the non-cooperative expression in Eq. (13) is recovered.

**II.C. Derivation of the kinetic equations for describing the binding process**

The kinetic equations describing the time variation in the number of ligands of type λ bound to the polymer is

$$\frac{dn_\lambda}{dt} = k_{b\lambda} n \bar{s}_\lambda - k_{r\lambda} n_\lambda , \tag{20}$$

where $k_{b\lambda}$ is the binding kinetic constant, $k_{r\lambda}$ is the release kinetic constant, and the number of gaps, which is $n + 1$ (counting all gaps with $g \geq 0$), is approximated by the total number of ligands

$$n = \sum_{i=1}^{k} n_i. \tag{21}$$

This kinetic equation, Eq. (20), extends for several binding modes the previous results in Refs. [26], [33]. Combining Eqs. (1), (20), and (21), we obtain the kinetic equations for the coverage

$$\frac{dc_\lambda}{dt} = \left( c_\lambda + m_\lambda \sum_{i \neq \lambda}^{k} \frac{c_i}{m_i} \right) k_{b\lambda} \bar{s}_\lambda - k_{r\lambda} c_\lambda . \tag{22}$$

## III. Two modes ligand binding

We consider that there are two different binding modes, labelled modes 1 and 2, with different binding rates, $k_{b1}$ and $k_{b2}$, and release rates, $k_{r1}$ and $k_{r2}$. Hence, the kinetic equations describing the time variation in the coverages are [using Eq. (22)]

$$\frac{dc_1}{dt} = \left( c_1 + m_1 \frac{c_2}{m_2} \right) k_{b1} \bar{s}_1 - k_{r1} c_1 , \tag{23}$$

$$\frac{dc_2}{dt} = \left( c_2 + m_2 \frac{c_1}{m_1} \right) k_{b2} \bar{s}_2 - k_{r2} c_2 . \tag{24}$$

Using Eqs. (3) and (4), we derive the following mathematical relations for the conditional probabilities

$$(ff) + (fb_1^1) + (fb_1^2) = 1 \tag{25}$$

$$\left( b_{m_1}^1 f \right) + \left( b_{m_1}^1 b_1^1 \right) + \left( b_{m_1}^1 b_1^2 \right) = 1 \tag{26}$$

$$\left( b_{m_2}^2 f \right) + \left( b_{m_2}^2 b_1^1 \right) + \left( b_{m_2}^2 b_1^2 \right) = 1 \tag{27}$$

To introduce various aspects of the problem, in Subsection III.A, we consider ligands that do not interact between them, the non-cooperative case. Next, in Subsection III.B, we derive results where ligand-ligand interaction is allowed between nearest neighbors, the cooperative case.

### III.A. Non-cooperative case

We will consider that both modes are non-cooperative, i.e., $\omega_{\alpha\beta} = 1$ in Eq. (10). We may write that $(b^\alpha_{m_\alpha} f) = (ff)$, which states that having selected the first residue of either type $b^\alpha_{m_\alpha}$, or of type $f$, the probability of the residue to its immediate right being of type $f$ is independent of the initial selection. Equivalently, $(b^\alpha_{m_\alpha} b^\beta_1) = (f b^\beta_1)$. Note that these equalities are only valid for the non-cooperative case.

We also derive the following mathematical relation, using Eqs. (6):

$$(1-c)(ff) + \frac{c_1}{m_1}(b^1_{m_1} f) + \frac{c_2}{m_2}(b^2_{m_2} f) = (1-c) \tag{28}$$

In that case, the average numbers of free binding sites [using Eq. (13)] in the polymer turn out to be

$$\bar{s}_1 = \frac{(ff)^{m_1}}{(1-(ff))}, \tag{29}$$

$$\bar{s}_2 = \frac{(ff)^{m_2}}{(1-(ff))}. \tag{30}$$

Moreover, Eq. (28) can be solved to obtain the following expression for $(ff)$,

$$(ff) = \frac{1 - c_1 - c_2}{1 - c_1 - c_2 + c_1/m_1 + c_2/m_2}. \tag{31}$$

Hence, the kinetic equations (23) and (24) for the coverage become

$$\frac{dc_1}{dt} = k_{b1} m_1 (1 - c_1 - c_2) \left(\frac{1 - c_1 - c_2}{1 - c_1 - c_2 + c_1/m_1 + c_2/m_2}\right)^{m_1 - 1} - k_{r1} c_1, \tag{32}$$

$$\frac{dc_2}{dt} = k_{b2} m_2 (1 - c_1 - c_2) \left(\frac{1 - c_1 - c_2}{1 - c_1 - c_2 + c_1/m_1 + c_2/m_2}\right)^{m_2 - 1} - k_{r2} c_2, \tag{33}$$

The equations (32) and (33) can be integrated numerically to obtain the time variation of the coverages. Fig. 2 shows examples of kinetics, and stresses that compensating a factor 2 difference in the size of the ligand requires nearly two orders of magnitude difference in the binding constant $k_b/k_r$.

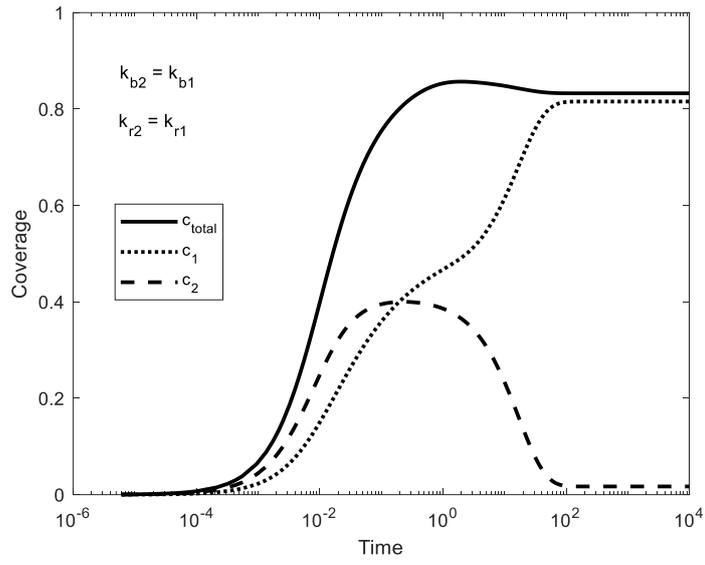

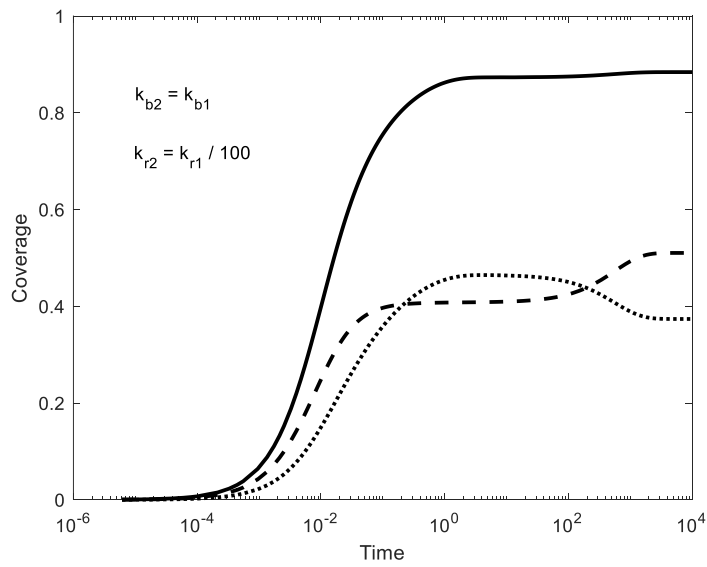

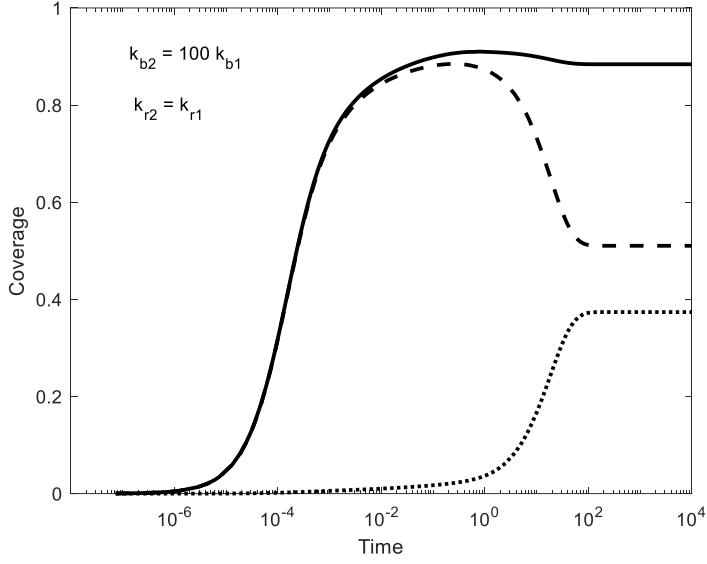

**FIG. 2: Polymer coverage of modes 1 and 2, and total coverage, as a function of time for non-cooperative system.** *Starting with a naked polymer, ligands binding to $m_1 = 30$ and $m_2 = 60$ sites. (Top) binding rates $k_{b1} = k_{b2} = 0.8$ s$^{-1}$, and release rates $k_{r1} = k_{r2} = 0.06$ s$^{-1}$. (Middle) binding rates $k_{b1} = k_{b2} = 0.8$ s$^{-1}$, and release rates $k_{r1} = 0.06$ s$^{-1}$ and $k_{r2} = 0.0006$ s$^{-1}$. (Bottom) binding rates $k_{b1} = 0.8$ s$^{-1}$ and $k_{b2} = 80$ s$^{-1}$, and release rates $k_{r1} = k_{r2} = 0.06$ s$^{-1}$.*

For large ligand size ($m_i \gg 1$), the following approximations are held true,

$$\left(\frac{1 - c_1 - c_2}{1 - c_1 - c_2 + c_1/m_1 + c_2/m_2}\right)^{m_1 - 1} = \exp\left(\frac{-c_1(1 + c_2 m_1/c_1 m_2)}{1 - c_1 - c_2}\right), \quad (34)$$

$$\left(\frac{1 - c_1 - c_2}{1 - c_1 - c_2 + c_1/m_1 + c_2/m_2}\right)^{m_2 - 1} = \exp\left(\frac{-c_2(1 + c_1 m_2/c_2 m_1)}{1 - c_1 - c_2}\right), \quad (35)$$

which provides a simplification that can be useful for applications.

### III.B. Cooperative case

In this subsection we aim to study the effect of cooperativity on competitive binding in a simple two ligands case. For the sake of simplicity, we assume that mode 1 is self-cooperative and mode 2 is non-cooperative. Thus, $\omega_{11} \neq 1$, $\omega_{22} = 1$, and $\omega_{12} = \omega_{21} = 1$. The equations for the cooperativity parameters become [using Eq. (10)]

$$\omega_{11} = \frac{(b^1_{m_1} b^1_1)(ff)}{(b^1_{m_1} f)(f b^1_1)} \quad (36)$$

$$\omega_{22} = \frac{(b_{m_2}^2 b_1^2)(ff)}{(b_{m_2}^2 f)(f b_1^2)} \tag{37}$$

$$\omega_{12} = \frac{(b_{m_1}^1 b_1^2)(ff)}{(b_{m_1}^1 f)(f b_1^2)} \tag{38}$$

$$\omega_{21} = \frac{(b_{m_2}^2 b_1^1)(ff)}{(b_{m_2}^2 f)(f b_1^1)} \tag{39}$$

The fact that the mode 1 is self-cooperative implies that $(b_{m_1}^1 f) \neq (ff)$ and $(b_{m_1}^1 b_1^1) \neq (f b_1^1)$. As the mode 2 is non-cooperative, [using Eq. (37)], it follows that $(b_{m_2}^2 f) = (ff)$. The fact that there is not cooperation between modes 1 and 2, [using Eqs. (38) and (39)] implies that $(b_{m_1}^1 b_1^2)(ff) = (b_{m_1}^1 f)(f b_1^2)$, and $(b_{m_2}^2 b_1^1)(ff) = (b_{m_2}^2 f)(f b_1^1)$. However, as the mode 1 is self-cooperative, the following additional inequalities hold $(b_{m_1}^1 b_1^2) \neq (f b_1^2)$, and $(b_{m_2}^2 b_1^1) \neq (f b_1^1)$.

For mode 1, the average number of free doubly contiguous binding sites [using Eq. (15) becomes

$$(\bar{s}_{dc})^1 = \left[ \frac{c_1}{c} (b_{m_1}^1 f) \omega_{11} + \frac{c_2}{c} \cdot (ff) \right] (ff)^{m_1 - 1} [(f b_1^1) \omega_{11} + (f b_1^2)], \tag{40}$$

the average number of free singly contiguous binding sites [using Eq. (17)] becomes

$$(\bar{s}_{sc})^1 = \left[ \frac{c_1}{c} \cdot (b_{m_1}^1 f)(f b_1^1) 2 \omega_{11} + \frac{c_1}{c} (b_{m_1}^1 f)(f b_1^2) \cdot (\omega_{11} + 1) + \frac{c_2}{c} (b_{m_2}^2 f)(f b_1^1) \right.$$
$$\left. \cdot (\omega_{11} + 1) + \frac{c_2}{c} (b_{m_2}^2 f)(f b_1^2) 2 \right] \frac{(ff)^{m_1}}{1 - (ff)}, \tag{41}$$

and the average number of isolated binding sites [using Eq. (18)] is

$$(\bar{s}_i)^1 = \left[ \frac{c_1}{c} (b_{m_1}^1 f) + \frac{c_2}{c} (b_{m_2}^2 f) \right] \frac{(ff)^{m_1 + 1}}{1 - (ff)}. \tag{42}$$

Finally, the average number of free binding sites of type 1 in the polymer is [using Eq. (19)]

$$\bar{s}_1 = (\bar{s}_{dc})^1 + (\bar{s}_{sc})^1 + (\bar{s}_i)^1 \tag{43}$$

For mode 2, the average number of free binding sites [using Eq. (13)] turns out to be

$$\bar{s}_2 = \left[ \frac{c_1}{c} (b_{m_1}^1 f) + \frac{c_2}{c} (b_{m_2}^2 f) \right] \frac{(ff)^{m_2 - 1}}{1 - (ff)}. \tag{44}$$

Next, we solved numerically Eqs. (25)-(27) and (36)-(39) to obtain nine different conditional probabilities, by using typical values of SSB binding to ssDNA. Then, Eqs. (23) and (24) can be integrated numerically to obtain the time variation of the coverages. The equilibrium coverage can be obtained numerically from Eqs. (23) and (24) imposing $dc_1/dt = 0; dc_2/dt = 0$. See Fig. 3.

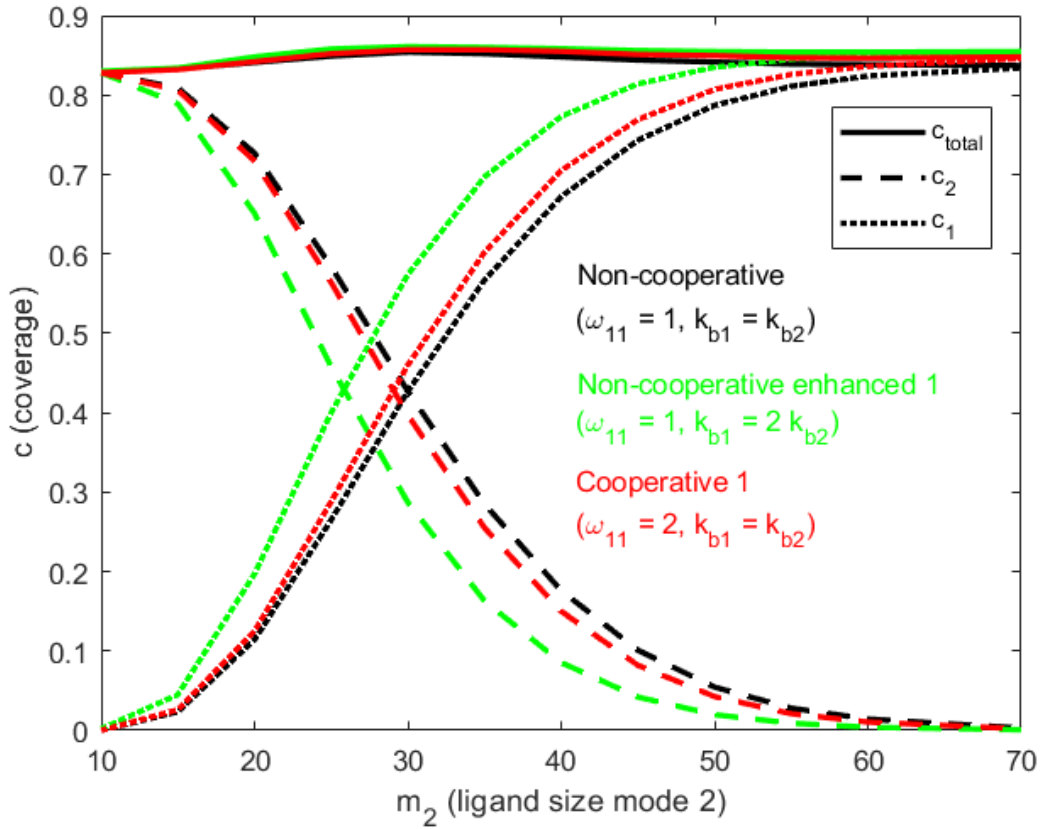

*FIG. 3: Comparison of equilibrium coverage of modes 1 and 2, and total coverage, as a function of mode 2 size. Model parameters:* $m_1 = 30$, $k_{r1} = k_{r2} = 0.06\ s^{-1}$; (black lines) Non-cooperative: $\omega_{11} = 1$, $k_{b1} = k_{b2} = 1\ s^{-1}$; (green lines) Non-cooperative enhanced of mode 1: $\omega_{11} = 1$, $k_{b1} = 2\ s^{-1}$, $k_{b2} = 1\ s^{-1}$; (red lines) Cooperative mode 1: $\omega_{11} = 2$, $k_{b1} = k_{b2} = 1\ s^{-1}$.

## **IV. Conclusions**

McGhee and von Hippel derived an insightful and thorough theoretical work describing the binding of large ligands to a long polymer [27]. They obtained the equilibrium coverage through the detailed computation of the possible binding sites for the ligand. Within this framework, we derived the kinetic equations for multiple ligands competing to bind to a linear polymer. These kinetic equations complement the previous equilibrium results [34]–[39]. The derived kinetic equations have been obtained doing an appropriate counting of the binding sites, following the McGhee-von Hippel counting procedure [27]. This appropriate counting is required for an appropriate account of the exclusion effect of bound ligands [26]. The derived kinetic equations provide a general framework to study competitive binding kinetics.

Our analysis of the two-ligand binding kinetics reveals that the size of the ligand (number of monomers occluded) is the dominant factor in competitive binding. Revealing that smaller ligands are more effective in covering or passivating the polymer. This result has a purely entropic origin. Thus, the polymer can host more ligands of smaller size, resulting in a larger number of polymer-ligand states. According to Boltzmann equation, this leads to a higher entropy of the system, meaning that the polymer interacting with smaller ligands will have a smaller free energy in comparison to the polymer interacting with larger ligands. This favours the interaction of the polymer with the smaller ligand [40]. Here, halving the ligand size has a larger effect than doubling the ligand binding affinity. Additionally, the effect of increasing the cooperativity is shown in this model to be smaller than the effect of increasing the affinity. (See Fig. 3).

The impact for the SSB mode competition is that smaller modes are more effective in covering and passivating the ssDNA. However, the detailed description of the SSB binding kinetics also requires the inclusion of the conversion rates between the modes involving details that are beyond the scope of the present study that require further analysis.

Here, we considered that the cooperativity is produced by an increase of binding affinity through interaction with a neighbouring bound ligand. Cooperativity may also emerge by a reduction of the release rate through interaction with a neighbouring bound ligand, or even by a combination of the two effects [33]. The competitive binding kinetic equations presented here can be generalized to the other cooperativity cases following the procedures indicated in Ref. [33] for single-ligand binding. Our motivation to study these systems is their application to the kinetics of SSB binding to single-stranded DNA (ssDNA), which can be cooperative or non-cooperative depending on the type of SSB protein and on its binding mode. The model presented here can be also useful to describe other cooperative binding processes, such the binding of ionic surfactants to the charged linear polyelectrolyte, the binding of lysozyme to chitosan, among others. In addition, the model can be applied for non-cooperative ligand binding, like the binding of oligolysine to nucleic acids, mono- and multivalent ions to polyelectrolytes, or other polypeptides to polynucleotide. Cooperativity has been addressed here for the case of nearest-neighbours interaction. However, salt-induced bridging might induce further reaching interactions, for example in SSB binding to ssDNA [41]. The formalism presented in Ref. [42] might allow the extension of the equilibrium results for interactions beyond nearest-neighbours.

The model presented here is valid to describe transitions between different equilibrium states of the ligands bound to the polymer, as at each coverage configuration the ligands are assumed to be distributed as in equilibrium. Relevant examples of transitions to apply the model are the binding dynamics of a ligand on a naked polymer, and the unbinding dynamics of a ligand from the polymer after the ligand is washed out of the media. However, it cannot describe the early dynamics from far from equilibrium ligand distributions on the polymer, where transient effects of this departure from equilibrium might modify the early dynamics.

In the present study we have not addressed the boundary effects (and related problem of the ligand position correlation), which has allowed to explain experimental observations on nucleosomes positioning. Refs. [43], [44] show that boundary conditions and excluded volume of large ligands lead to the formation of patterns in the ligand position distribution. Ref. [40] considers the same problem with interactions between nearby positions (as in our cooperative case), and ref. [42] provides equilibrium results for ligand with interaction beyond nearest neighbour position. These studies focus on the impact of boundary conditions on the equilibrium ligand distribution, particularly relevant for short polymers (DNA fragments). Our study instead addresses the kinetics and competitive equilibrium between competing modes on large polymers. Effects of the boundary conditions on the mode competition at equilibrium are also discussed on ref. [40]. Extension of the study to the computation of the ligand position distribution, and to the effects of boundary conditions to the kinetics would be also relevant for the applications to SSB and nucleosome ligand binding to DNA. Extensions to 2D of the model presented here could serve to describe the ligand binding kinetics to passivate an active surface.

**Acknowledgment**

This work was supported by the Spanish Ministry of Economy and Competitiveness and the European Regional Development Fund through grant RTI2018-095802-B-I00.

**A. Appendix: Derivation of the average number of free binding sites per polymer**

A gap consists of any consecutive $g$ free ligand binding sites (free lattice residues). For a gap $g$ residues long between adjacent bound ligands, the number of binding sites is $g - m_i + 1$ if $g \geq m_i$, but zero if $g < m_i$. The probability of a gap having $g$ free polymer sites long is denoted by $P_g$, and the probability of a gap having $g$ free polymer sites long between a bound ligand of type α and a bound ligand of type β, is denoted by $(P_g)_{\alpha\beta}$. The relation between both probabilities is given by

$$P_g = \sum_{\alpha=1}^{k}\sum_{\beta=1}^{k}(P_g)_{\alpha\beta}. \tag{A1}$$

Using combinatorial arguments, the probability of a gap having $g$ free polymer sites long between a bound ligand of type α and a bound ligand of type β is given by

$$(P_g)_{\alpha\beta} = (b_{m_\alpha}^\alpha f)(ff)^{g-1}(fb_1^\beta), \tag{A2}$$

where $(b_{m_\alpha}^\alpha f)$ is the probability, having selected the right end of a bound ligand of type α, that the polymer site to the immediate right is free; $(ff)$ is the probability, given a free site, that there is a second free site to the immediate right; and $(fb_1^\beta)$ is the probability, given a free site, that the left end of a bound ligand of type β lies to the immediate right.

### A.1. Non-cooperative case

When the binding is non-cooperative ($\omega_{lh} = 1$), the average number of free binding sites in the polymer, $\bar{s}_\lambda$, is given by

$$\bar{s}_\lambda = \sum_{\alpha=1}^{k}\sum_{\beta=1}^{k}\sum_{g=m_\lambda}^{N}(g - m_\lambda + 1)\frac{c_\alpha}{c}(P_g)_{\alpha\beta}. \tag{A3}$$

The factor $c_\alpha/c$ is the probability that the gap is preceded by a ligand of type $\alpha$. By letting $N$ go to infinity, we obtain that

$$\bar{s}_\lambda = \sum_{\alpha=1}^{k}\sum_{\beta=1}^{k}\sum_{g=m_\lambda}^{N}(g-m_\lambda+1)\frac{c_\alpha}{c}(P_g)_{\alpha\beta}=$$

$$=\left[\sum_{\alpha=1}^{k}\frac{c_\alpha}{c}(b_{m_\alpha}^\alpha f)\right]\left[\sum_{g=m_\lambda}^{\infty}(g-m_\lambda+1)(ff)^{g-1}\right]\left[\sum_{\beta=1}^{k}(fb_1^\beta)\right]=$$

$$=\left[\sum_{\alpha=1}^{k}\frac{c_\alpha}{c}(b_{m_\alpha}^\alpha f)\right]\frac{(ff)^{m_\lambda-1}}{(1-(ff))^2}\left[\sum_{\beta=1}^{k}(fb_1^\beta)\right]=$$

$$=\left[\sum_{\alpha=1}^{k}\frac{c_\alpha}{c}(b_{m_\alpha}^\alpha f)\right]\frac{(ff)^{m_\lambda-1}}{(1-(ff))^2}[1-(ff)]=\left[\sum_{\alpha=1}^{k}\frac{c_\alpha}{c}(b_{m_\alpha}^\alpha f)\right]\frac{(ff)^{m_\lambda-1}}{1-(ff)}, \quad (A4)$$

where Eq. (3), and the following mathematical derivation was used

$$\sum_{g=m_\lambda}^{\infty}(g-m_\lambda+1)\cdot(ff)^{g-1}=(ff)^{m_\lambda-1}+2(ff)^{m_\lambda}+\cdots=$$

$$=(ff)^{m_\lambda-2}\sum_{n=1}^{\infty}n(ff)^n=(ff)^{m_\lambda-2}(ff)\frac{d}{d(ff)}\left[\sum_{n=1}^{\infty}(ff)^n\right]=$$

$$=(ff)^{m_\lambda-1}\frac{d}{d(ff)}\left[(ff)\sum_{n=0}^{\infty}(ff)^n\right]=$$

$$=(ff)^{m_\lambda-1}\frac{d}{d(ff)}\left[\frac{(ff)}{1-(ff)}\right]=\frac{(ff)^{m_\lambda-1}}{(1-(ff))^2}. \quad (A5)$$

Furthermore, by using $(b_{m_\alpha}^\alpha f)=(ff)$ for the non-cooperative case and $c=\sum_{\alpha=1}^{k}c_\alpha$, we obtain that

$$\bar{s}_\lambda=\left[\sum_{\alpha=1}^{k}\frac{c_\alpha}{c}(b_{m_\alpha}^\alpha f)\right]\frac{(ff)^{m_\lambda-1}}{1-(ff)}=(ff)\frac{(ff)^{m_\lambda-1}}{1-(ff)}=\frac{(ff)^{m_\lambda}}{1-(ff)}. \quad (A6)$$

### A.2. Cooperative case

When the binding is cooperative ($\omega_{lh}\neq 1$), there are three distinct types of ligand binding sites. For $g<m_i$, there are no free binding sites. For $g=m_i$, there is one doubly binding site per gap, thus

$$(\bar{s}_{dc})_{\alpha\beta}^\lambda=(P_{m_\lambda})_{\alpha\beta}=\frac{c_\alpha}{c}(b_{m_\alpha}^\alpha f)(ff)^{m_\lambda-1}(fb_1^\beta), \quad (A7)$$

by summing over the different combinations weighted by the cooperativities,

$$(\bar{s}_{dc})^\lambda = \sum_{\alpha=1}^{k}\sum_{\beta=1}^{k}(\bar{s}_{dc})_{\alpha\beta}^\lambda \omega_{\lambda\alpha}\omega_{\lambda\beta} =$$

$$= \left[\sum_{\alpha=1}^{k}\frac{c_\alpha}{c}(b_{m_\alpha}^\alpha f)\omega_{\lambda\alpha}\right](ff)^{m_\lambda-1}\left[\sum_{\beta=1}^{k}(fb_1^\beta)\omega_{\lambda\beta}\right]. \quad (A8)$$

The sums are weighted by the cooperativities to properly account for their impact in the kinetic equations. A cooperativity greater than one implies an increased affinity of the ligand for this site. These cooperativity effects are then already accounted during the binding site counting, which is thus an effective site counting in the cooperative case.

For $g \geq m_i + 1$, there are two singly contiguous binding site per gap, thus the sum turns out to be

$$(\bar{s}_{sc})_{\alpha\beta}^\lambda = 2\sum_{g=m_\lambda+1}^{\infty}(P_g)_{\alpha\beta} = 2\sum_{g=m_\lambda+1}^{\infty}\frac{c_\alpha}{c}(b_{m_\alpha}^\alpha f)(ff)^{g-1}(fb_1^\beta) =$$

$$= 2\frac{c_\alpha}{c}(b_{m_\alpha}^\alpha f)\frac{(ff)^{m_\lambda}}{1-(ff)}(fb_1^\beta), \quad (A9)$$

In the previous derivation, we used the following mathematical expression

$$\sum_{g=m_\lambda+1}^{\infty}(ff)^{g-1} = (ff)^{m_\lambda}\sum_{n=1}^{\infty}(ff)^{n-1} = \frac{(ff)^{m_\lambda}}{1-(ff)} \quad (A10)$$

Next summing over the different possible ligand boundaries and weighting with the cooperativities

$$(\bar{s}_{sc})^\lambda = \sum_{\alpha=1}^{k}\sum_{\beta=1}^{k}(\bar{s}_{sc})_{\alpha\beta}^\lambda \frac{\omega_{\alpha\lambda}+\omega_{\lambda\beta}}{2}$$

$$= \sum_{\alpha=1}^{k}\sum_{\beta=1}^{k}\frac{c_\alpha}{c}(b_{m_\alpha}^\alpha f)\frac{(ff)^{m_\lambda}}{1-(ff)}(fb_1^\beta)(\omega_{\alpha\lambda}+\omega_{\lambda\beta}). \quad (A11)$$

For $g \geq m_i + 2$, there are $(g - m_i - 1)$ isolated binding sites per gap, thus the sum in the polymer becomes

$$(\bar{s}_i)^\lambda = \sum_{\alpha=1}^{k}\sum_{\beta=1}^{k}\sum_{g=m_\lambda+2}^{\infty}(g-m_\lambda-1)(P_g)_{\alpha\beta} =$$

$$= \sum_{\alpha=1}^{k}\sum_{\beta=1}^{k}\sum_{g=m_\lambda+2}^{\infty}(g-m_\lambda-1)\frac{c_\alpha}{c}(b_{m_\alpha}^\alpha f)(ff)^{g-1}(fb_1^\beta) =$$

$$= \left[\sum_{\alpha=1}^{k}\frac{c_\alpha}{c}(b^\alpha_{m_\alpha}f)\right]\left[\sum_{g=m_\lambda+2}^{\infty}(g-m_\lambda-1)(ff)^{g-1}\right]\left[\sum_{\beta=1}^{k}(fb^\beta_1)\right]=$$

$$=\left[\sum_{\alpha=1}^{k}\frac{c_\alpha}{c}(b^\alpha_{m_\alpha}f)\right]\frac{(ff)^{m_\lambda+1}}{(1-(ff))^2}\left[\sum_{\beta=1}^{k}(fb^\beta_1)\right], \quad (A12)$$

where $\bar{s}_i$ is the average number of isolated binding sites. In the derivation of Eq. (A12), the following mathematical expression was used

$$\sum_{g=m_\lambda+2}^{\infty}(g-m_\lambda-1)(ff)^{g-1} = (ff)^{m_\lambda+1}+2(ff)^{m_\lambda+2}+\cdots =$$

$$= (ff)^{m_\lambda}\sum_{n=1}^{\infty}n(ff)^n = (ff)^{m_\lambda}(ff)\frac{d}{d(ff)}\left[\sum_{n=1}^{\infty}(ff)^n\right]=$$

$$= (ff)^{m_\lambda+1}\frac{d}{d(ff)}\left[(ff)\sum_{n=0}^{\infty}(ff)^n\right]=$$

$$= (ff)^{m_\lambda+1}\frac{d}{d(ff)}\left[\frac{(ff)}{1-(ff)}\right] = \frac{(ff)^{m_\lambda+1}}{(1-(ff))^2}. \quad (A13)$$

Finally, by using Eqs. (A8), (A11), and (A12), the average number of free binding sites in the polymer is written as:

$$\bar{s}_\lambda = (\bar{s}_{dc})^\lambda + (\bar{s}_{sc})^\lambda + (\bar{s}_i)^\lambda =$$

$$=\sum_{\alpha=1}^{k}\sum_{\beta=1}^{k}\frac{c_\alpha}{c}(b^\alpha_{m_\alpha}f)(fb^\beta_1)(ff)^{m_\lambda-1}\cdot\left[\omega_{\alpha\lambda}\omega_{\lambda\beta}+(\omega_{\alpha\lambda}+\omega_{\lambda\beta})\frac{(ff)}{1-(ff)}+\frac{(ff)^2}{(1-(ff))^2}\right]=$$

$$=\sum_{\alpha=1}^{k}\sum_{\beta=1}^{k}\frac{c_\alpha}{c}(b^\alpha_{m_\alpha}f)(fb^\beta_1)(ff)^{m_\lambda-1}\left[\omega_{\alpha\lambda}+\frac{(ff)}{1-(ff)}\right]\left[\omega_{\lambda\beta}+\frac{(ff)}{1-(ff)}\right]=$$

$$=\left\{\sum_{\alpha=1}^{k}\frac{c_\alpha}{c}(b^\alpha_{m_\alpha}f)\left[\omega_{\alpha\lambda}+\frac{(ff)}{1-(ff)}\right]\right\}(ff)^{m_\lambda-1}\left\{\sum_{\beta=1}^{k}(fb^\beta_1)\left[\omega_{\lambda\beta}+\frac{(ff)}{1-(ff)}\right]\right\}. \quad (A14)$$

Consistently, the non-cooperative expression in Eq. (A4) is recovered from Eq. (A14) when $\omega_{lh}=1$.

**B. Derivation of the average number of free binding sites in a system with two binding modes**

**B.1. Non-cooperative case**

We will consider that both modes are non-cooperative, i.e., $\omega_{\alpha\beta} = 1$. In that case, the average number of free binding sites [using Eq. (A6)] of modes 1 and 2 turn out to be

$$\bar{s}_1 = \frac{(ff)^{m_1}}{1-(ff)}, \qquad (A15)$$

$$\bar{s}_2 = \frac{(ff)^{m_2}}{1-(ff)}. \qquad (A16)$$

**B.2. Cooperative case**

For the sake of simplicity, we will assume that mode 1 is self-cooperative and mode 2 is non-cooperative, i.e., $\omega_{11} \neq 1$, $\omega_{12} = \omega_{21} = \omega_{22} = 1$. Using Eqn. (A8), the average number of free doubly contiguous binding sites of type 1 in the polymer is

$$(\bar{s}_{dc})^1 = \left[\frac{c_1}{c}(b^1_{m_1}f)\omega_{11} + \frac{c_2}{c}(b^2_{m_2}f)\right](ff)^{m_1-1}[(fb^1_1)\omega_{11} + (fb^2_1)], \qquad (A17)$$

where the relation $(b^2_{m_2}f) = (ff)$ can be used because the mode 2 is non-cooperative. Using Eq. (A11) the average number of free singly contiguous binding sites of type 1 in the polymer is

$$(\bar{s}_{sc})^1 = \left[\frac{c_1}{c}(b^1_{m_1}f)(fb^1_1)2\omega_{11} + \frac{c_1}{c}(b^1_{m_1}f)(fb^2_1)(\omega_{11}+1)\right.$$
$$\left. + \frac{c_2}{c}(b^2_{m_2}f)(fb^1_1)(\omega_{11}+1) + \frac{c_2}{c}(b^2_{m_2}f)(fb^2_1)2\right]\frac{(ff)^{m_1}}{1-(ff)}, \qquad (A18)$$

Using Eq. (A12) the average number of free isolated binding sites of type 1 in the polymer is

$$(\bar{s}_i)^1 = \left[\frac{c_1}{c}(b^1_{m_1}f) + \frac{c_2}{c}(b^2_{m_2}f)\right]\frac{(ff)^{m_1+1}}{(1-(ff))^2}[(fb^1_1) + (fb^2_1)] =$$
$$\left[\frac{c_1}{c}(b^1_{m_1}f) + \frac{c_2}{c}(b^2_{m_2}f)\right]\frac{(ff)^{m_1+1}}{(1-(ff))^2}[1-(ff)] =$$
$$= \left[\frac{c_1}{c}(b^1_{m_1}f) + \frac{c_2}{c}(b^2_{m_2}f)\right]\frac{(ff)^{m_1+1}}{1-(ff)}. \qquad (A19)$$

Using Eqn. (A6), we derive the following relation for the average number of free binding sites of mode 2 in the polymer

$$\bar{s}_2 = \left[\frac{c_1}{c}(b^1_{m_1}f) + \frac{c_2}{c}(b^2_{m_2}f)\right]\frac{(ff)^{m_2-1}}{1-(ff)}. \qquad (A20)$$